\documentstyle[12pt]{article}

\newcommand{\be}{\begin{equation}}
\newcommand{\ee}{\end{equation}}

\newcommand{\ra}{\rightarrow}
\newcommand{\vp}{\varphi}

\newcommand{\prt}{\partial}

\newcommand{\om}{\omega}

\newcommand{\ep}{\varepsilon}

\newcommand{\br}{{\bf r}}

\newcommand{\bk}{{\bf k}}

\newcommand{\bs}{{\bf s}}

\begin{document}

\begin{center}
{\Large{\bf Stratification of Moving Multicomponent Bose-Einstein
Condensates} \\ [5mm]

V.I. Yukalov$^1$ and E.P. Yukalova$^2$} \\ [5mm]

{\it
$^1$Bogolubov Laboratory of Theoretical Physics \\
Joint Institute for Nuclear Research, Dubna 141980, Russia\\ [3mm]

$^2$Department of Computational Physics \\
Laboratory of Information Technologies \\
Joint Institute for Nuclear Research, Dubna 141980, Russia}

\end{center}

\vskip 2cm

\begin{abstract}

A mixture of the multicomponent Bose-Einstein condensate is
considered, where each component moves with its own velocity. As
a result of the relative motion, the mixture stratifies when the
relative velocity reaches a critical value. Stability conditions
for a binary moving mixture are derived and the critical velocity
is found.

\end{abstract}

\newpage

Ultracold atomic gases, forming Bose-Einstein condensates, are
nowadays an active field of research (see reviews [1--5]). Dilute
Bose gas can be trapped and cooled down to temperatures when
practically all atoms are in Bose-Einstein condensate. One can
create mixtures of several Bose-condensed species. The stability
of such mixtures depends on the mutual atomic interactions. The
condition of stability for binary mixtures was, first, found by
Nepomnyashchy [6], considering the collective excitation spectrum
in the Bogolubov and long-wave approximations. When this stability
condition is not valid, the mixture is not stable and stratifies
onto separate components. The analysis of Nepomnyashchy [6] was
based on the Green functions approach for a superfluid mixture
with broken gauge symmetry. As was shown later [7], the same
stability condition holds true for a fluid mixture with conserved
gauge symmetry.

Recently, it has become feasible to create moving wave packets
of Bose-Einstein condensates, which could collide and move
through each other [8--12]. Therefore, it is interesting to study
the interplay between the possibility of different Bose components
to mix and the influence on their stability of their relative motion 
through each other. Such a relative motion of two fluids could 
change the stability condition of mixing [6,7]. Bose-Einstein
condensates, formed in modern traps, are made of dilute gases,
for which the Gross-Pitaevskii equation provides a very good
description [1--5]. Actually, this equation is an exact equation
for the coherent system at zero temperature [13]. The aim of the
present communication is to analyze a mixture of Bose-Einstein
condensates, whose components move through each other, under
conditions of low temperature, when the system is described by
the Gross-Pitaevskii equation. This situation corresponds to
current experiments with Bose-Einstein condensates of dilute
gases.

An atomic gas, of $N$ atoms in volume $V$, is dilute when its
density $\rho\equiv N/V$ and the $s$-wave scattering length $a$
are connected by the inequality $\rho a^3\ll 1$. This can be
generalized to a mixture of species enumerated by the index
$i=1,2,\ldots$ as the inequality $\sqrt{\rho_i\rho_j}a_{ij}^3\ll 1$,
in which $a_{ii}\equiv a_i$  and $a_{ij}$ are the corresponding
scattering lengths, and $\rho_i\equiv N_i/V$. The Gross-Pitaevskii
equation for a multicomponent mixture is
\be
\label{1}
i\hbar \; \frac{\prt}{\prt t}\; \eta_j(\br,t) =
\hat H_j[\eta]\; \eta_j(\br,t) \; ,
\ee
where $\eta_i(\br,t)$ is the coherent field of the $i$-component;
in the nonlinear Hamiltonian
\be
\label{2}
\hat H_i[\eta] = \frac{\hat{\bf p}^2}{2m_i} + U_i +
\sum_j A_{ij}|\eta_j|^2
\ee
the first term is the kinetic-energy operator, with $\hat{\bf p}
\equiv -i\hbar\vec\nabla$, and $m_i$ being atomic mass, $U_i$ is an
external (trapping) potential, and the third term describes contact
atomic interactions with
$$
A_{ij} \equiv 2\pi\hbar^2 a_{ij}\;
\frac{m_i+m_j}{m_im_j} \; .
$$
The normalization condition for each component reads
$$
||\eta_i||^2 \equiv \int |\eta_i(\br,t)|^2\; d\br =
N_i \; .
$$

Elementary collective excitations of the system can be obtained
by linearizing the Gross-Pitaevskii equation (1) around a given
stationary solution. Following here this way, we should keep in mind
that our aim here is to study a mixture of moving components. Assume
that each component moves with a fixed velocity $\bs_i$. Then taking
account of this motion can be done by means of the Galilean
transformation replacing $\eta_j$ by $\eta_j\exp\{\frac{i}{\hbar}m_j
\bs_j\cdot\br\}$. To find collective excitations in the moving mixture,
we write the coherent field $\eta_j$ in the form
\be
\label{3}
\eta_j(\br,t) =\left [ \vp_j(\br) + u_j(\br) e^{-i\om t} +
v_j^*(\br) e^{i\om t} \right ] \exp\left \{ \frac{i}{\hbar}
\left (m_j\; \bs_j\cdot \br - E_j t\right )\right \} \; ,
\ee
where $\vp_j(\br)$ satisfies the stationary equation
\be
\label{4}
\left (\hat H_i[\vp] +\hat{\bf p}\cdot\bs_i + \frac{m_is_i^2}{2}
\right )\; \vp_i(\br) = E_i\vp_i(\br) \; ,
\ee
with the normalization $||\vp_i||^2=N_i$. Substituting Eq. (3) into
the Gross-Pitaevskii equation (1) and linearizing it with respect
to the functions $u_j(\br)$ and $v_j(\br)$, we obtain
$$
\left (\hbar\om - \hat{\bf p}\cdot\bs_i -\hat H[\vp] - \frac{m_is_i^2}{2}
+ E_i\right ) u_i - \vp_i \sum_j A_{ij} \left ( \vp_j^* u_j +
\vp_j v_j \right ) = 0 \; ,
$$
\be
\label{5}
\left (\hbar\om - \hat{\bf p}\cdot\bs_i + \hat H[\vp] + \frac{m_is_i^2}{2}
- E_i\right ) v_i + \vp_i^* \sum_j A_{ij} \left ( \vp_j^* u_j +
\vp_j v_j \right ) = 0 \; .
\ee
This is a generalization of the Bogolubov - de-Gennes equations to the
case of a mixture of moving components.

Being interested in collective excitations above the ground state of
trapped atoms, we keep in mind that the wave functions, defined by
Eq. (4), are real-valued. Then we can resort to the local-density
approximation, setting $\vp_i(\br)\simeq\sqrt{\rho_i(\br)}$ and
assuming that $u_i$ and $v_i$ are proportional to $e^{i\bk\cdot\br}$.
Supposing that atomic interactions are sufficiently strong, so that
mean potential energies are essentially larger that mean kinetic
energies, we may omit the terms containing the spatial differentiation
of $\rho_i(\br)$. In this approximation,
\be
\label{6}
\left ( \hat H_i[\vp] + \frac{m_is_i^2}{2}\right ) u_i =
\left ( E_i + \frac{\hbar^2k^2}{2m_i}\right ) u_i \; .
\ee
Then Eqs. (5) reduce to
$$
\hbar\left ( \om -\bk\cdot\bs_i -\; \frac{\hbar k^2}{2m_i}\right ) u_i -
\sum_j A_{ij}\sqrt{\rho_i\rho_j} (u_j + v_j) = 0 \; ,
$$
\be
\label{7}
\hbar\left ( \om -\bk\cdot\bs_i + \frac{\hbar k^2}{2m_i}\right ) v_i +
\sum_j A_{ij}\sqrt{\rho_i\rho_j} (u_j + v_j) = 0 \; .
\ee
The latter equations possess a nontrivial solution if the related
determinant is zero. To write down the resulting equation in a compact
form, we introduce the notation
\be
\label{8}
\om_i^2 \equiv c_i^2 k^2 + \left ( \frac{\hbar k^2}{2m_i}
\right )^2 \; , \qquad c_i^2 \equiv \frac{\rho_i}{m_i}\; A_i \; ,
\ee
where $A_i\equiv A_{ii}$. If there would be no interactions between
different species, then each component would have the Bogolubov spectrum
$\om_i(k)$ given by Eq. (8). Note that the existence of the Bogolubov
spectrum for a single-component Bose-Einstein condensate of trapped
atoms is well confirmed by experiment [14]. Define also
\be
\label{9}
\om_{ij} \equiv c_{ij} k \; , \qquad c_{ij}^2 \equiv \left (
\frac{\rho_i\rho_j}{m_im_j}\right )^{1/2} A_{ij} \; .
\ee
And let us consider the case of a binary mixture. The spectrum of
collective excitations, following from Eq. (7), is given by the
equation
\be
\label{10}
\left [ (\om -\ep_1)^2 - \om_1^2 \right ] \left [ (\om -\ep_2)^2 -
\om_2^2\right ] = \om_{12}^4 \; ,
\ee
in which $\ep_i\equiv\bk\cdot\bs_i$.

It is convenient to connect the system of coordinates with one of the
components, say with the first one, so that $\bs_1=0$. Then $\bs_2\equiv
\bs$ is the velocity of the second component with respect to the first
component. In that case, the spectral equation (10) yields
\be
\label{11}
\om^4-2\ep\om^3 -\left ( \om_1^2 +\om_2^2 -\ep^2\right )
\om^2 + 2\ep\om_1^2 \om + \om_1^2 \left ( \om_2^2 -\ep^2\right ) -
\om_{12}^4 = 0 \; ,
\ee
where
\be
\label{12}
\ep\equiv \bk\cdot \bs = ks \cos\vartheta \; .
\ee
Equation (11) has four solutions for $\om=\om(\bk)$. But only two of
them can be stable, being positively defined for all $\vartheta\in[0,2\pi]$.
The necessary condition for Eq. (11) to possess two positive solutions
for arbitrary $\ep\in[-ks,ks]$, by Descartes theorem, is
\be
\label{13}
\om_1^2 \om_2^2 -\om_{12}^4 > \om_1^2 k^2 s^2 \; .
\ee
In the long-wave limit, when $k\ra 0$, we have the {\it stability
condition}
\be
\label{14}
c_1^2 c_2^2 - c_{12}^4 > c_1^2 s^2 \; .
\ee
This shows that, when the immovable mixture is stable, satisfying the
inequality $c_1^2c_2^2>c_{12}^4$, it will retain its stability only up
to the {\it critical velocity}
\be
\label{15}
s_c = \frac{1}{c_1}\sqrt{c_1^2 c_2^2 - c_{12}^4} \; .
\ee
Reaching this relative velocity of one component through another,
the binary mixture becomes unstable and stratifies onto two spatially
separated components. The stratification starts inside the cone of the
angle
\be
\label{16}
\vartheta_c ={\rm arc cos}\; \frac{s_c}{s} \; .
\ee
The stability condition (14), with Eqs. (8) and (9), transforms to
\be
\label{17}
A_1 A_2 - A_{12}^2 > \frac{m_2}{\rho_2}\; A_1 s^2 \; .
\ee
Respectively, the critical velocity (15) can be written as
\be
\label{18}
s_c = \sqrt{\frac{\rho_2(A_1A_2-A_{12}^2)}{m_2A_1} } \; .
\ee
Taking into account the relation between $A_{ij}$ and the scattering length
$a_{ij}$, the stability condition (17) can be presented in the form
\be
\label{19}
\frac{a_1a_2}{a_{12}^2} > \frac{(m_1+m_2)^2}{4m_1m_2} +
\frac{m_2^2a_1 s^2}{4\pi\hbar^2\rho_2a_{12}^2} \; .
\ee
Finally, for colliding wave packets with nonuniform densities
$\rho_i(\br)$, the stratification would occur not everywhere in space,
where the moving packets overlap, but only there, where the inequality
\be
\label{20}
\rho_2(\br) > \frac{m_2A_1s^2}{A_1A_2-A_{12}^2}
\ee
becomes invalid, provided that $A_1A_2>A_{12}^2$.

The stability conditions for the mixture of components with relative
motion have been obtained here by considering the stability of the
spectrum of collective excitations. This is what is called the dynamic
stability [2]. There exists as well the notion of thermodynamic
stability, when one compares the corresponding free energies, or,
at zero temperature, the related internal energies. For immovable
mixtures, these two types of stability result in equivalent stability
conditions. However, for the mixtures of moving components, these
two kinds of stability are not equivalent. To show the latter, we may,
for simplicity, employ the uniform approximation, writing the internal
energy of the mixture as
\be
\label{21}
E_{mix} = \frac{1}{2V}\left ( N_1^2 A_1 + N_2^2 A_2 + 2N_1 N_2 A_{12}
\right ) + K \; ,
\ee
where $V=V_1+V_2$ and the last term is the kinetic energy
\be
\label{22}
K =\frac{1}{2}\left ( N_1 m_1 s_1^2 + N_2 m_2 s_2^2 \right ) \; .
\ee
If the components are spatially separated, that is, stratified, their
total internal energy is
\be
\label{23}
E_{str} =  \frac{N_1^2A_1}{2V_1} + \frac{N_2^2 A_2}{2V_2} + K \; .
\ee
Because of the additivity of kinetic energy, the comparison of $E_{mix}$
and $E_{str}$ does not involve the velocities of the moving components,
so that the thermodynamic stability condition $E_{mix}<E_{str}$ does not
depend on the relative motion of these components. Moreover, employing
the thermodynamic stability condition, one commonly invokes the condition
of mechanical equilibrium, according to which the pressures of the
stratified components,
$$
p_i \equiv -\; \frac{\prt E_{str}}{\prt V_i} =
\frac{N_i^2A_i}{2V_i} \; ,
$$
are equal, so that $p_1=p_2$. The latter yields the equality
$$
\frac{N_1^2}{V_1^2}\; A_1  = \frac{N_2^2}{V_2^2}\; A_2 \; ,
$$
from where, recalling that $V=V_1+V_2$, we have the relations
$$
\frac{V}{V_1} =  1 + \sqrt{\frac{N_2^2A_2}{N_1^2A_1}} \; , \qquad
\frac{V}{V_2} = 1 + \sqrt{\frac{N_1^2 A_1}{N_2^2 A_2}} \; .
$$
This makes it possible to present Eq. (23) as
\be
\label{24}
E_{str} = \frac{1}{2V} \left ( N_1^2 A_1 + N_2^2 A_2 +
2N_1 N_2 \sqrt{A_1 A_2}\right ) + K \; .
\ee
Now, comparing Eqs. (21) and (24), one notices that $E_{mix}<E_{str}$ if
and only if $\sqrt{A_1A_2}>A_{12}$. However, the usage of the condition
of mechanical equilibrium for moving separated components does not look
appropriate. The thermodynamic stability condition can be applied only to
immovable mixtures.

The method of studying the stability of mixtures with a relative motion
of their components by means of the dynamic stability condition, requiring
the positive definiteness for the spectrum of collective excitations, is
quite general and can be applied to arbitrary mixtures. In particular, this
can also be an atomic-molecular mixture. Such mixtures are formed by means
of the Feshbach resonance technique [2,15]. Two Bose atoms can form a
quasi-molecule. Then, it is possible to create a two-component mixture
consisting of atoms themselves and of the molecules formed by these atoms.
When both atoms and molecules are in the Bose-condensed state, they again
can be described by the system of the Gross-Pitaevskii equations for the
corresponding coherent fields. Thus, if the interaction amplitude for the 
reaction "atom+atom $\leftrightarrow$ molecule" is denoted by $B$, then 
the related Gross-Pitaevskii equations are 
\be
\label{25}
i\hbar\; \frac{\prt}{\prt t}\; \eta_1 = \hat H_1[\eta]\eta_1 +
B\eta_1^*\eta_2 \; , \qquad
i\hbar\; \frac{\prt}{\prt t}\; \eta_2 = \hat H_2[\eta]\eta_2 +
\frac{1}{2}\; B\eta_1^2 \; ,
\ee
where $\hat H_i[\eta]$ are the same Hamiltonians (2). When atoms as well
as molecules are immovable, their mixture is stable under the condition
$$
\sqrt{A_1A_2} > A_{12} + \frac{B}{\sqrt{\rho_2}} \; ,
$$
where $\rho_2\equiv N_2/V$. But if one is able to organize the relative
motion of the atomic and molecular components with respect to each other,
then, to find the stability condition for such a moving mixture, one has
to follow the same procedure as described above. That is, the form (3)
of the coherent fields has to be substituted into Eqs. (25), after which
the spectrum of collective excitations can be obtained. Requiring the
positiveness of this spectrum yields the stability condition under the
relative motion of components.

Concluding, if an immovable mixture of Bose-condensed species is
stable, it can stratify if the components move with respect to each
other. The stratification occurs when the relative velocity of motion
reaches a critical value. The conditions of stability and, respectively,
stratification can be derived by linearizing the Gross-Pitaevskii
equations for the components, with the inclusion of the Galilean
transformation for the coherent fields, as is done in Eq. (3).

\newpage

\begin{center}
{\bf Corresponding author}
\end{center}

Prof. V.I. Yukalov                                            

\vskip 2mm

Bogolubov Laboratory of Theoretical Physics

Joint Institute for Nuclear Research

Dubna 141980, Russia

\vskip 3mm

{\bf E-mail}: yukalov@thsun1.jinr.ru

{\bf Tel}: 7 (096) 216-3947 (office); 7 (096) 213-3824 (home)

\vskip 2mm

{\bf Fax}: 7 (096) 216-5084

\end{document}